# Hybrid Semantic Search: Unveiling User Intent Beyond Keywords

Aman Ahluwalia, Bishwajit Sutradhar, Karishma Ghosh, Indrapal Yadav,
Arpan Sheetal, Prashant Patil

Innoplexus Consulting Services Pvt Ltd, Pune, Maharashtra, India

**Abstract**
This paper addresses the limitations of traditional keyword-based search in understanding user intent and introduces a novel hybrid search approach that leverages the strengths of non-semantic search engines, Large Language Models (LLMs), and embedding models. The proposed system integrates keyword matching, semantic vector embeddings, and LLM-generated structured queries to deliver highly relevant and contextually appropriate search results. By combining these complementary methods, the hybrid approach effectively captures both explicit and implicit user intent.The paper further explores techniques to optimize query execution for faster response times and demonstrates the effectiveness of this hybrid search model in producing comprehensive and accurate search outcomes.

## Introduction

Information retrieval has traditionally relied on keyword-based search engines [1]. While these methods identify documents containing the queried terms, they often fail to capture the user's true intent, leading to irrelevant results. Semantic search represents a paradigm shift in information retrieval by transcending the limitations of traditional keyword-based approaches using non-semantic models like BM25 [1]. At its core, semantic search hinges on two crucial components. The first, the search function, acts similarly to traditional search engines [1] by identifying and ranking documents relevant to a user's query within a vast collection of information (corpus). However, semantic search goes beyond this basic functionality with its second component: semantic understanding. This is where Transformers come into play, allowing the system to delve deeper than keyword matching. The introduction of Transformers [2], a powerful deep learning architecture, revolutionized search by enabling semantic understanding. Through Transformers, semantic search can grasp the underlying meaning and intent behind both the user's query and the retrieved documents, ultimately leading to a more accurate and relevant search experience. The adoption of semantic search by major search engines has significantly improved user experience. Despite the development of encoder-decoder transformers and decoder-only transformers, the related fields of semantic search and language embeddings are still dominated by small encoder transformers [3]. Recent advances in semantic search approaches have seen a rise in the use of decoder-only transformers to extract semantically meaningful sentence embeddings [4].

Keyword-based search engines, while prevalent, exhibit limitations in capturing the intricacies of user intent and handling synonyms or paraphrases [5][6]. On the other hand, sentence embedding techniques have

emerged as a powerful tool for semantic search, enabling retrieval based on semantic similarity rather than exact keyword matches. However, these methods can be susceptible to noise introduced by irrelevant words within the query or documents, potentially hindering retrieval accuracy. Furthermore, sentence embeddings can face challenges when dealing with very short or long queries, as short queries might lack sufficient context for accurate representation, and long queries can be computationally expensive.

This work proposes a hybrid search model that merges the strengths of both keyword-based search and sentence embedding techniques to address these limitations. This synergistic approach aims to achieve a balance between retrieval precision and semantic understanding. This hybrid strategy strives to achieve a more comprehensive and user-centric search experience by combining the strengths of both established and emerging retrieval techniques.

**The Hybrid Approach:**
The hybrid search model introduced in this paper merges three complementary search methodologies to optimize document retrieval. Firstly, LLM-based structured query generation leverages the language model's ability to understand and interpret user intent, transforming ambiguous queries into precise search terms. Secondly, traditional keyword search is employed to identify documents containing exact matches to the query terms, ensuring high recall. Thirdly, semantic search, utilizing embedding models, captures the underlying meaning and context of the query, expanding the search space to include semantically related documents. By combining these distinct approaches, the model comprehensively addresses various query types and retrieval challenges. The resulting search results are then meticulously reranked to prioritize documents that most closely align with the user's information need. Here's a detailed breakdown of each step:

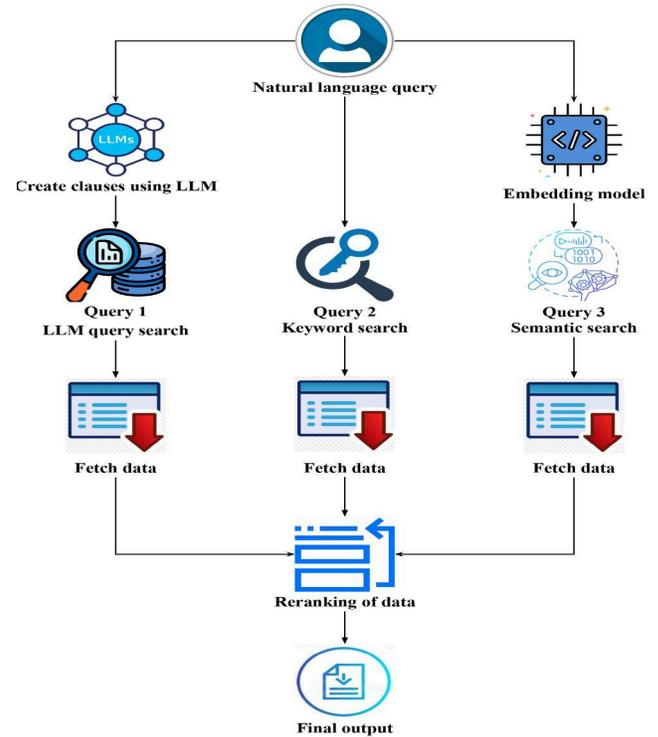

*Fig 1:* Flowchart of the hybrid approach

**1. LLM-Based Structured Query Generation:**
The first step in the hybrid search model involves the use of a Large Language Model (LLM) to convert the user's natural language query into a structured query format. A Large Language Model (LLM) is instrumental in bridging the gap between natural language queries and structured database queries. It serves as a semantic interpreter, translating ambiguous user intent into precise search terms. This is achieved through several sub-steps:

**a. Understanding the User Query:** When a user inputs a query, such as "lung cancer in India," the LLM first parses the query to understand its underlying components. The model identifies key elements within the query, such as the "indication" (lung cancer) and the "location" (India). This understanding is rooted in the LLM's deep training on vast amounts of text data, enabling it to recognize patterns, relationships, and context within the query. The LLM assigns semantic types to extracted entities. For instance, "EGFR" would be classified as a gene and "lung cancer" as a disease category. This step is crucial for mapping entities to the appropriate database fields.

**b. Extracting Relevant Entities and Attributes:** After understanding the query, the LLM extracts specific entities (e.g., diseases, locations, dates) and attributes (e.g., age groups, conditions) that are pertinent to the search. For example, in the query "atopic dermatitis in adults in India since 2022," the LLM identifies "atopic dermatitis" as the indication, "adults" as the age group, "India" as the location, and "2022" as the starting year.

**c. Generating the Structured Query:** Using the extracted entities and attributes, the LLM constructs a structured query in a format that can be easily processed by the search engine. For the example "atopic dermatitis in adults in India since 2022," the LLM might generate a structured query such as:

$\{ "indication": "atopic\ dermatitis",$
$\ "age": \{ "\$gt": 18 \},$
$\ "country": "India",$
$\ "year": \{ "\$gte": 2022 \} \}$

This structured query allows the search engine to perform a more focused retrieval, targeting specific aspects of the user's intent.

**2. Keyword Search:** The second component of the hybrid model is the keyword search mechanism. Unlike the structured query generation, which relies on semantic understanding, the keyword search method focuses on directly matching the user's query terms with the text in the documents.

**a. Tokenization of the User Query:** The first step in keyword search is to break down the user's query into individual tokens (words or phrases). For instance, the query "lung cancer in India" would be tokenized into "lung," "cancer," and "India."

**b. Query Expansion (Optional):** Depending on the implementation, the system may perform query expansion, where synonyms or related terms are added to the original query to increase the chances of retrieving relevant documents. For example, "lung cancer" might be expanded to include "lung carcinoma."

**c. Matching Tokens with Document Text:** The search engine then scans the corpus of documents to find matches for the query tokens. Documents that contain the highest number of matching tokens, particularly in close proximity to each other, are considered more relevant. The search results are ranked based on the frequency and distribution of these matches within the documents.

**d. Scoring and Ranking:** Each document is assigned a relevance score based on the number and quality of keyword matches. Documents with higher scores are ranked higher in the search results. The simplicity and speed of keyword search make it an essential component of the hybrid model, particularly for queries that are well-defined by specific terms.

**3. Semantic Search Using Vectors**: The third component of the hybrid model is semantic search, which relies on vector representations of both the query and the documents to find semantically similar content, even if the exact keywords are not present.

**a. Generating Query Embeddings:** When a user submits a query, it is first converted into an embedding vector. An embedding vector is a numerical representation of the query in a high-dimensional space. This vector captures the semantic meaning of the query by considering the relationships between words based on the context in which they appear.

**b. Document Embeddings:** Each document in the corpus is also pre-processed to generate its own embedding vector. These document vectors are stored in a database, enabling quick retrieval during the search process. The document embeddings capture the semantic content of the entire document or specific sections of it.

**c. Cosine Similarity Search:** The core of semantic search is the comparison between the query vector and the document vectors. This comparison is done using cosine similarity, a metric that measures the cosine of the angle between two vectors. Vectors that are closer together (i.e., have a smaller angle between them) are considered more similar in meaning. The search engine retrieves documents whose vectors have the highest cosine similarity to the query vector.

**d. Ranking Based on Semantic Relevance:** The documents retrieved through semantic search are ranked based on their similarity scores. This allows the search engine to surface documents that are contextually relevant, even if they do not contain the exact words used in the query.

**4. Reranking of Results:** After the three search mechanisms—LLM-based structured query search, keyword search, and semantic search—have retrieved their respective sets of documents, the final step is to combine and rerank these results.

**a. Aggregation of Results:** The documents retrieved by the three methods are aggregated into a single list. Each document in this list is associated with a relevance score from its respective search method.

**b. Reciprocal Rank Fusion (RRF) Algorithm:** The reranking process uses the Reciprocal Rank Fusion (RRF) algorithm, which is designed to merge ranked lists from different search engines or methods. The RRF algorithm gives a higher rank to documents that are ranked highly across multiple lists, ensuring that the most relevant documents are prioritized. In this context, it recalibrates the ranking scores of documents retrieved by the LLM-based structured query search, the keyword search, and the semantic search. The RRF formula for a document $d$ is given by:

$$RRF(d) = \sum_{k=1}^{n} \frac{1}{rank_k(d) + c}$$

Where $n$ is the number of ranked lists (three in our case), $rank_k(d)$ is the rank of document $d$ in the $k^{th}$ list, and $c$ is a constant to control the influence of lower-ranked documents. This formula ensures that documents ranked highly across multiple lists are given precedence in the final output.

**c. Final Output:** The final output is a reranked list of documents that represent the best combination of keyword matches, structured query results, and semantic relevance. This reranked list is presented to the user, offering a comprehensive and contextually accurate set of search results.

**5. Tailoring the Search Model to User Needs:** To accommodate varying search requirements, the proposed hybrid model offers flexibility in its execution. For scenarios demanding high throughput, the LLM-based query generation component can be bypassed, streamlining the search process. This optimization prioritizes speed over comprehensive semantic understanding. Conversely, when retrieval accuracy and precision are paramount, all three search mechanisms are employed to deliver highly relevant results. By dynamically adjusting the model's configuration based on specific use cases, practitioners can effectively balance search performance and quality.

**Observations:**

In the course of applying the hybrid search model to various datasets, several key observations were made:

**1. Improved Retrieval Accuracy:** The combination of structured queries, keyword matching, and semantic embeddings led to more accurate retrieval results, particularly in scenarios where traditional keyword-based search would struggle to understand user intent.

**2. Robustness Across Query Types:** The hybrid model demonstrated robustness across different types of queries, including short, long, and ambiguous queries. While keyword search excelled in matching specific terms, the LLM and semantic searches provided the necessary context to improve overall relevance.

**3. Performance Considerations:** Although the hybrid approach introduces additional computational steps, the benefits in terms of retrieval accuracy outweigh the costs. Furthermore, the reranking process ensures that the final results are both comprehensive and efficient.

**Conclusions:**

This paper presents a novel hybrid search model that integrates LLM-based query structuring, keyword search, and semantic vector search to enhance document retrieval accuracy. By addressing the limitations of traditional keyword-based search engines and capitalizing on the strengths of semantic understanding, the proposed model offers a more comprehensive approach to information retrieval.

The hybrid search model demonstrated its effectiveness in various scenarios, consistently delivering relevant and contextually appropriate search results. The use of the Reciprocal Rank Fusion algorithm for reranking further optimized the final output, ensuring that the most pertinent documents are prioritized.

Future work could explore the application of this hybrid model to more specialized domains, as well as the integration of additional machine learning techniques for further refinement. Additionally, optimizing the computational efficiency of the hybrid model remains an area for ongoing research.

**Acknowledgements:**

We would like to express our sincere gratitude to the entire Innoplexus Consulting Services Pvt Ltd, India team for their invaluable contributions to this research.

In particular, we are indebted to Vatsal Agarwal and Sudhir Mansukh for their mentorship and guidance throughout this project. The encouragement and insightful feedback from Pankaj Jain, Rachana Nagar, Gayatri Bhawariya, Lakshmi Geeta, Sailaja Ayyagari, Deepa Tatkare and Gayatri Bhawariya were instrumental in accomplishing this project. Special thanks to Naveen Dara, Pranjal Verma, Aasif Shaikh and Madhab Doley for their technical assistance.